\documentclass[11pt, a4paper]{article}
\usepackage{natbib}
\usepackage{graphicx}
\usepackage{enumerate}
\usepackage[top=2.7cm, bottom= 2.4cm, left=2.5cm, right=2.5cm]{geometry}
\usepackage{footmisc}
\usepackage{setspace}
\usepackage{fancyhdr}
\usepackage{amsmath,amssymb,mathrsfs}
\usepackage{appendix}
\usepackage[normal]{caption2}
\usepackage[figuresleft]{rotating}
\usepackage{threeparttable}
\usepackage{hyperref}
\usepackage{longtable}
\usepackage{pdflscape}
\usepackage{float}
\usepackage{bigfoot}
\usepackage{color}
\usepackage{appendix}
\usepackage[british,UKenglish,USenglish,english,american]{babel}
\providecommand*\email[1]{\href{mailto:#1}{#1}}

\hypersetup{
colorlinks = true, 
urlcolor   = blue,
linkcolor  = blue,
citecolor  = blue
}

\global\dimen\footins = 3cm

\usepackage{authblk}

\usepackage{tabularx, blindtext}
\newcommand{\argmin}{\operatornamewithlimits{argmin}}

\title{\textbf{A nowcasting approach to generate timely estimates of Mexican economic activity: An application to the period of COVID-19}}

\author[1]{Francisco Corona\thanks{Corresponding author: \email{franciscoj.corona@inegi.org.mx}. Please, if you require to quote this working progress, request it to the corresponding author. The views expressed here are those of the authors and do not reflect those of INEGI.}}
\author[2]{Graciela Gonz\'alez-Far\'ias}
\author[1]{Jes\'us L\'opez-P\'erez}
\affil[1]{Instituto Nacional de Estad\'istica y Geograf\'ia}
\affil[2]{Centro de Investigaci\'on en Matem\'aticas A.C.}

\date{This version: November 6, 2020}

\bibpunct{(}{)}{,}{a}{,}{,}
\include{macros1}

\bibpunct{(}{)}{,}{a}{,}{,}
 \def\beqlb{\begin{equation}}\def\eeqlb{\end{equation}}
 \def\beqnn{\begin{eqnarray*}}\def\eeqnn{\end{eqnarray*}}
 
 \def\<{\langle}\def\>{\rangle}
 
\begin{document}
\maketitle

\begin{abstract}
In this paper, we present a new approach based on dynamic factor models (DFMs) to perform nowcasts for the percentage annual variation of the Mexican Global Economic Activity Indicator (IGAE in Spanish). The procedure consists of the following steps: i) build a timely and correlated database by using economic and financial time series and real-time variables such as social mobility and significant topics extracted by Google Trends; ii) estimate the common factors using the two-step methodology of \cite{Dozetal2011}; iii) use the common factors in univariate time-series models for test data; and iv) according to the best results obtained in the previous step, combine the statistically equal better nowcasts (Diebold-Mariano test) to generate the current nowcasts. We obtain timely and accurate nowcasts for the IGAE, including those for the current phase of drastic drops in the economy related to COVID-19 sanitary measures. Additionally, the approach allows us to disentangle the key variables in the DFM by estimating the confidence interval for both the factor loadings and the factor estimates. This approach can be used in official statistics to obtain preliminary estimates for IGAE up to 50 days before the official results.
\end{abstract}


\newpage








\section{Introduction}
Currently, the large amount of economic and financial time series collected over several years by official statistical agencies allows researchers to implement statistical and econometric methodologies to generate accurate models to understand any macroeconomic phenomenon. One of the most important events to anticipate is the movement of the gross domestic product (GDP) because doing so allows policy to be carried out with more certainty, according to the expected scenario. For instance, if an economic contraction is foreseeable, businesses can adjust their investment or expansion plans, governments can apply countercyclical policy, and consumers can adjust their spending patterns.

As new economic and financial information is released, the forecasts for a certain period are constantly also being updated; thus, different GDP estimations arise. In this sense, a new, unexpected event can drastically affect predictions in the short term; consequently, it might be necessary to use not only economic and financial information but also nontraditional and high-frequency indicators, such as news, search topics extracted from the Internet, social networks, etc. The seminal work of \cite{varian2014big} is an obligatory reference for the inclusion of high-frequency information by economists, and \cite{buono2018big} is also an important reference to characterize the types of nontraditional data and see the econometric methods usually employed to extract information from these data.

Thus, the term ``nowcast", or real-time estimation, is relevant because we can use a rich variety of information to model, from a multivariate point of view, macroeconomic and financial events, plus specific incidents that can affect the dynamics of GDP in the short run. Econometrically and statistically, these facts are related to the literature on large dynamic factor models (DFMs) because a large amount of time series is useful to estimate underlying common factors. First introduced in economics by \cite{Geweke1977} and \cite{SargentSims1977}, DFMs have recently become very attractive in practice given the current requirements of dealing with large datasets of time series using high-dimensional DFM; see, for example, \cite{BreitungEickmeir2006}, \cite{BaiNg2008}, \cite{StockWatson2011}, \cite{BreitungChoi2013} and \cite{BaiWang2016} for reviews of the existing literature.

An open question in the literature on large DFMs is whether a large number of series is adequate for a particular forecasting objective. In that sense, preselecting variables has proven to reduce the error prediction with respect to using the complete dataset \cite{BoivinNg2005}; that is, not always by using a large set of variables, we can obtain closer factor estimates with respect to when we use fewer variables, especially under finite sample performance \cite{PoncelaRuiz2015b}. Even when the number of time series is moderate, approximately 15, we can accurately estimate the simulated common factors, as shown by \cite{Coronaetal2020} in a Monte Carlo analysis. The latter also corroborates that the \cite{Dozetal2011} two-step (2SM) factor extraction method performs better than other approaches available in the literature above all when the data are nonstationary.

DFM methodology has already been used to nowcast or predict the Mexican economy. \cite{Coronaetal2017}, one of the first works in this line, estimated common trends in a large and nonstationary DFM to predict the Global Economic Activity Indicator (IGAE in Spanish) two steps ahead and concluded that the error prediction was reduced with respect to some benchmarking univariate and multivariate time-series models. \cite{Caruso2018} focuses on international indicators, mainly for the US economy, to show that its nowcasts of quarterly GDP outperform the predictions obtained by professional forecasters. Recently, \cite{Galvez2020} concluded that bridge equations perform better than DFM and static principal components (PCs) when making the nowcasts of quarterly GDP. An important work related with timely GDP estimation is \cite{guerreroetal2013} where, based on vector autoregression (VAR) models, they generate rapid GDP estimates (and its three grand economic activities) with a delay of up to 15 days from the end of the reference quarter, while the official GDP takes around 52 days after the quarter closes. This work is the main reference to INEGI's ``\textit{Estimaci\'on Oportuna del PIB Trimestral}."\footnote{\url{https://www.inegi.org.mx/temas/pibo/}}

Although prior studies are empirically relevant for the case of Mexico, our analysis goes beyond including nontraditional information to capture more drastic frictions that occur in the very short run, one or two months. We identify that previous works focus on traditional information, which limits their capacity to predict the recent historical declines attributed to COVID-19 and the associated economic closures since March 2020. Our approach maximizes the structural explanation of the already relevant macroeconomic and financial time series with the timeliness of other high-frequency variables commonly used in big data analysis.

In this tradition, this work estimates a flexible and trained DFM to verify the assumptions that guarantee the consistency of the component estimation from a statistical point of view. That is, we use previous knowledge and attempt to fill in the identified gaps by focusing on the Mexican case in the following ways: i) build a timely and correlated database by using traditional economic and financial time series and real-time nontraditional information, determining the latter relevant variables with least absolute selection and shrinkage operator (LASSO) regression, a method of variable selection; ii) estimate the common factors using the two-step methodology of \cite{Dozetal2011}; iii) train univariate time series models with the DFM's common factors to select the best nowcasts; iv) determine the confidence intervals for both the factor loadings and the factor itself to analyze the importance of each variable and the uncertainty attributed to the estimation; and iv) combine the statistically equal better nowcasts to generate the current estimates.

In practice, we consider the benefits of this paper to be opportunity and openness. First, given the timely availability of the information that our approach uses, we can generate nowcasts of the IGAE up to 50 days before the official data release; thus, our approach becomes an alternative to obtaining IGAE's preliminary estimates, which are very important in official statistics. Second, this paper illustrates the empirical strategy to generate IGAE nowcasts step-by-step to practitioners, so any user can replicate the results for other time series. Third, and very important, the nowcasting approach allows to known which variables are the most relevant in the nowcasts, consequently, we emphasize in the structural explanation of our results.

The remainder of this paper is structured as follows. The next section, \ref{sect2}, summarizes the Mexican economy evolution in the era of COVID-19. Section \ref{sect3} presents the methodology considered to generate the nowcasts. Section \ref{sect4} describes the data and the descriptive analysis. Section \ref{sect5} contains the empirical results. Finally, Section \ref{sect6} concludes the paper.

\section{The Mexican economy amid the COVID-19 pandemic} \label{sect2}

The first six months of the COVID-19 pandemic (until September 2020) has had severe impacts on the Mexican economy. The first case of coronavirus in Mexico was documented on February 27, 2020. Despite government efforts to cope with the effects of the obligatory halt of economic activity, GDP in the second quarter plummeted with a historic 18.7\% yearly contraction. Moreover, the pandemic accelerated economic stagnation that had begun to show signs of amelioration, following three quarters of negative growth of 0.5, 0.8 and 2.1\% since the third quarter of 2019. However, starting in 2020, the actual values were not foreseen by national and international institutions such as private and central banks. 
For example, the November 2019 Organisation for Economic Co-operation and Development Economic Outlook estimated the real GDP variation for 2020 at 1.29\%, while the June 2020 report updated it to -8.6\%, a difference of 9.8\%  in absolute terms. Moreover, even when the Mexican Central Bank expected barely zero economic growth for 2020, placing its November 2019 outlook between -0.2\% and 0.2\%, it did not anticipate such a contraction as has seen so far this year.

Between January 2019 and February 2020, before the COVID-19 outbreak started in Mexico, the annual growth of IGAE\footnote{The IGAE is the monthly proxy variable for Mexican GDP, which covers approximately 95\% of the total economy. Its publication is available two months after the reference month (\url{https://www.inegi.org.mx/temas/igae/}).} already showed signs of slowing and fluctuated around -1.75 and 0.76\%, and since May 2019, the economy exhibited nine consecutive months of negative growth. Broken down by sector and using IGAE, the economy suffered devastating consequences in the secondary and tertiary sectors. Overall, the pandemic brought about -19.7, -21.6 and -14.5\% contractions in total economic activity for April, May and June of 2020, respectively.

The industrial sector registered the deepest contractions, reducing its activity in April and May by -30.1 and -29.6\%, respectively, in annual terms, mainly driven by the closure of manufacturing and construction operations, which were considered nonessential businesses, following a slight recovery in June, -17.5\%, when an important number of activities, including automobile manufacturing, resumed but remained at low activity levels. The services sector also suffered from lockdown measures, falling by -15.9, -19 and -13.6\% in the three months of the second quarter, respectively, especially due to transportation, retail, lodging and food preparation, mainly due to the decrease in tourist activity, although restaurants and airports were not closed. The primary sector showed signs of resilience and even grew in April and May 2020, by 1.4 and 2.7\%, and only shrank in June by -1.5\% on an annual basis.

The great confinement in Mexico, which officially lasted from March 23 to May 31 (named ``Jornada Nacional de Sana Distancia"), had severe consequences for the components of the aggregate demand: consumption, investment and foreign trade suffered consequences. Consumption had been on a deteriorating path since September 2019, and in May 2020, the last month for which data are available, it exhibited a -23.5\% plunge compared to the same period of 2019. Similarly, investment, which peaked in June 2018, continued to deteriorate and registered a drop of -38.4\% in May 2020 on a year-over-year basis. Regarding international trade, exports began to abate in August 2019, hit a record low in May 2020, and despite a slight recovery in June, the yearly variation in July 2020 was still -8.8\% below its 2019 level. Similarly, imports registered a maximum in November 2018, and despite improvements in May 2020, the yearly variation as of July 2020 was still -26.3\% under its 2019 level. 

Prices and employment, to round out description of the Mexican economy, also suffered the ravages of the pandemic. Prices, unlike during other periods of economic instability in Mexico, do not seem to be into an inflationary spiral; in fact, the inflation rate in July 2020 compared to the previous year was 3.6\%, and the central bank expects it will hover around 3\% for the next 12 to 24 months. Additionally, different job-related statistics also reveal an underutilization of the labor force. For example, IMSS-affiliated workers, who account for approximately 90\% of the formal sector, suffered 1.3 million in job losses from the peak in November 2019 to July 2020. Similarly, the underemployment rate, an indicator of part-time employment, increased over twelve months from 7.6\% to 20.1\% in June 2020. In addition, the labor force participation rate showed a sharp decline in the first months of the social distancing policies, implying that 12 million people were dropped from the economy's active workforce thanks to COVID-19. Thus, the unemployment rate, people actively looking for a remunerated job, registered an annual increase of 1.32\% in June 2020 to stand at 5.5\%.

The literature on the effects of the pandemic on the economy has grown rapidly; see, for instance, \textit{Covid Economics} from the Center for Economic and Policy Research and numerous working papers from the National Bureau of Economic Research. For the case of the Mexican economy, the works of \cite{Campos2020laborcovid} who analyze online job advertisements in times of COVID-19 and \cite{Lustig2020covidlockdowns} who conducts simulations to project poverty figures across different population sectors by using survey's microdata, stand out. Along the same lines, the journal \textit{Econom\'iaUNAM} dedicated its number 51 of volume 17 in its entirety to study the impacts in Mexico of the pandemic, covering a wide range of issues related mainly to health economics \citep{vanegas2020desafios, kershenobich2020fortalezas}, labor  economics \citep{samaniego2020covid} , inequality \citep{alberro2020pandemia}, poverty \citep{fernandez2020pandemia}  and public policy \citep{sanchez2020mexico, moreno17pandemia}. None of these related to short-term forecasting of the economic activity. 

The closest paper to ours is \cite{meza2020forecasting}, who projects the economic impact of COVID-19 for twelve variables, including IGAE, based on a Susceptible-Infectious-Recovered epidemic model and a novel method to handle a sequence of extreme observations when estimating a VAR model \citep{lenza2020estimate}. To make the forecasts, \cite{meza2020forecasting} first estimates the shocks that hit the economy since  March 2020, and then produce four forecasts considering a path for the pandemic or not, and if so then considers three scenarios. Opposite to this work, the forecast horizon focuses in the mid term, June 2020 to February 2023, rather than ours in the short term, one or two months ahead.

\section{Methodology} \label{sect3}
This section describes how we employ DFM to generate the nowcasts of the IGAE. First, we describe how LASSO regression is used as a variable selection method to select among various Google Trends topics. Then, we report how the stationary DFM shrinks the complete dataset in the 2SM strategy to obtain the estimated factor loadings and common factors and in the \cite{Onatski2010} procedure to detect the number of common factors. Finally, we describe the nowcasting approach.

\subsection{LASSO regression}

LASSO regression was introduced by \cite{Tibshirani1996} as a new method of estimation in linear models by minimizing the residual sum of the squares (RSS) subject to the sum of the absolute value of the coefficients being less than a constant. In this sense, LASSO regression is related to ridge regression, but the former focuses on determining the tuning parameter, $\lambda$, that controls the regularization effect; consequently, we can have better predictions than ordinary least squares (OLS) in a variety of scenarios, depending on its choice.

Let $W_t = (w_{1t}, \dots, w_{Kt})'$ be a $K \times 1$ vector of stationary and standardized variables. Consider the following penalized RSS:

\begin{equation} \label{eq1}
 \min_{RSS} = (y - W \beta)'(y - W \beta) \quad \mbox{s.t} \quad f(\beta) \leq c,
\end{equation}
where $y = (y_1, \dots, y_T)'$ is a $T \times 1$ vector, $\beta = (\beta_1, \dots \beta_K)'$ is a $K \times 1$ vector, $W = (W1, \dots, W_T)'$ is a $T \times K$ matrix and $c \geq 0$ is a tuning parameter that controls the shrinkage of the estimates.

If $f(b) =  \sum_{j=1}^K \beta_j^2$, the ridge solution is $\widehat{\beta}^{Ridge}_{\lambda} = (W'W - \lambda I_p)^{-1} W'y$. In practice, this solution never sets coefficients to exactly zero; therefore, ridge regression cannot perform as a variable selection method in linear models, although its prediction ability is better than OLS.

\cite{Tibshirani1996} considers a penalty function as $f(\beta) = \sum_{j=1}^K |\beta_j| \leq c$; in this case, the solution of (\ref{eq4}) is not closed, and it is obtained by convex optimization techniques. The LASSO solution has the following implications: i) when $\lambda \rightarrow 0$, we obtain solutions similar to OLS, and ii) when $\lambda \rightarrow \infty$, $\widehat{\beta}^{LASSO}_{\lambda} \rightarrow 0.$ Therefore, LASSO regression can perform as a variable selection method in linear models. Consequently, if $\lambda$ is large, more coefficients tend to zero, selecting the variables that minimize the error prediction.

In macroeconomic applications, \cite{ApriglianoBencivelli2013} use LASSO regression to select the relevant economic and financial variables in a large data set with the goal of estimating a new Italian coincident indicator.

\subsection{Dynamic Factor Model}

We consider a stationary DFM where the observations, $X_t$, are generated by the following process:

\begin{equation}
X_{t}= PF_{t} + \varepsilon_{t},  \label{eq2}
\end{equation}
\begin{equation}
\Phi(L) F_{t} = \eta _{t},  \label{eq3}
\end{equation}
\begin{equation}
\Gamma(L) \varepsilon_{t} =  a_{t},  \label{eq4}
\end{equation}
where $X_{t}=(x_{1t},\dots ,x_{Nt})^{\prime }$ and $\varepsilon
_{t}=(\varepsilon _{1t},\dots ,\varepsilon _{Nt})^{\prime }$ are $N\times 1$ vectors of the variables and idiosyncratic noises observed at time $t$. The common factors, $F_{t}=(F_{1t},\dots ,F_{rt})^{\prime }$, and the factor disturbances, $\eta _{t}=(\eta_{1t},\dots ,\eta_{rt})^{\prime }$, are $r\times 1$ vectors, with $r$ $(r < N)$ being the number of static common factors, which is assumed to be known. The $N\times 1$ vector of idiosyncratic disturbances, $a_{t}$, is distributed independently of the factor disturbances, $\eta _{t}$, for all leads and lags, denoted by $L$, where $LX_t = X_{t-1}$. Furthermore, $\eta _{t}$ and $a_{t}$, are assumed to be Gaussian white noises with positive definite covariance matrices $\Sigma _{\eta }=\text{diag}(\sigma _{\eta _{1}}^{2},\dots,\sigma _{\eta _{r}}^{2})$ and $\Sigma _{a},$ respectively. $P=(p_{1},\dots ,p_{N})^{\prime}$, is the $N\times r$ matrix of factor loadings, where, $p_{i}=(p_{i1},\dots ,p_{ir})^{\prime}$ is an $r \times 1$ vector. Finally, $\Phi(L) = I - \sum_{i = 1}^k \Phi L^i$ and $\Gamma = I - \sum_{j = 1}^s \Gamma L^j$, where $\Phi$ and $\Gamma$ are $r\times r$ and $N\times N$ matrices containing the VAR parameters of the factors and idiosyncratic components with $k$ and $s$ orders, respectively. For simplicity, we assume that the number of dynamic factors, $r_1$, is equal to $r$.

Alternative representations in the stationary case are given by \cite{Dozetal2011, Dozetal2012}, who assume that $r$ can be different from $r_1$. Additionally, when $r = r_1$, \cite{BaiNg2004}, \cite{Choi2016}, and \cite{Coronaetal2020} also assume possible nonstationarity in the idiosyncratic noises. \cite{Barigozzietal2016, Barigozzietal2017} assume possible nonstationarity in $F_t$, $\varepsilon_t$ and $r \neq r_1$.

The DFM in equations (\ref{eq2}) to (\ref{eq4}) is not
identified. As we noted in the Introduction, the factor extraction used in this work is the 2SM; consequently, in the first step, we estimate the common factors by using PCs to solve the identification problem and uniquely define the
factors; we impose the restrictions $P^{\prime }P/N=I_{r}$
and $F^{\prime }F$ being diagonal, where $F=(F_{1},\dots ,F_{T})$ is $r\times T$. For a review of restrictions in the context of PC factor extraction, see \cite{BaiNg2013}.

\subsubsection{Two-step method for factor extraction} \label{sec2sm}
\cite{Giannoneetal2008} popularized the usage of 2SM factor extraction to estimate the common factors by using monthly information with the goal of generating the nowcasts of quarterly GDP. However,  \cite{Dozetal2011} proved the statistical consistency of the estimated common factor using 2SM. In the first step, PC factor extraction consistently estimates the static common factors without assuming any particular distribution, allowing weak serial and cross-sectional correlation in the idiosyncratic noises; see, for example, \cite{Bai2003}. In the second step, we model the dynamics of the common factors via the Kalman smoother, allowing idiosyncratic heteroskedasticity, a situation that occurs frequently in practice. In a finite sample study, \cite{Coronaetal2020} show that with the 2SM of \cite{Dozetal2011} based on PC and Kalman smoothing, we can obtain closer estimates of the common factors under several data generating processes that can occur in empirical analysis, such as heteroskedasticity and serial and cross-sectional correlation in idiosyncratic noises. Additionally, following \cite{Giannoneetal2008}, this method is useful when the objective is nowcasting given the flexibility to estimate common factors when all variables are not updated at the same time.

The 2SM procedure is implemented according to the following steps:

\begin{enumerate}
\item Set $\hat{P}$ as $\sqrt{N}$ times the $r$ largest eigenvalues of $X'X$, where $X = (X_1, \dots, X_T)'$ is a $T \times N$ matrix. By regressing $X$ on $\hat{P}$ and using the identifiability restrictions, obtain $\hat{F} = X \hat{P}/N$ and $\hat{\varepsilon} = X-\hat{F}'\hat{P}'.$ Compute the asymptotic confidence intervals for both factor loadings and common factors as proposed by \cite{Bai2003}.

\item Set the estimated covariance matrix of the idiosyncratic errors as $\hat{\Psi}=\text{diag}\left( \hat{\Sigma}_{\varepsilon}\right)$, where the diagonal of $\hat{\Psi}$ includes the variances of each variable of $X$; hence, $\hat{\sigma}^2_i$ for $i = 1, \dots, N.$

\item  Estimate a VAR(k) model by OLS to the estimated common factors, $\hat{F}$, and compute their estimated autoregressive coefficients as the VAR(1) model, denoted by $\hat{\Phi}$. Assuming that $f_{0}\sim N(0,\Sigma _{f})$, the unconditional
covariance matrix of the factors can be estimated as $\text{vec}\left( \hat{\Sigma}_{f}\right) =\left( I_{r^{2}}-\hat{\Phi}\otimes \hat{\Phi}\right) ^{-1}\text{vec}\left( \hat{\Sigma}_{\eta }\right)$, where $\hat{\Sigma}_{\eta }= \hat{\eta}'\hat{\eta}/T$.

\item Write DFM in equations (\ref{eq2}) to (\ref{eq4}) in state-space form, and with the system matrices substituted by $\hat{P}$, $\hat{\Psi}$, $\hat{\Phi}$, $\hat{\Sigma}_{\eta }$ and $\hat{\Sigma}_{f},$ use the Kalman smoother to obtain an updated estimation of the factors denoted by $\tilde{F}$.
\end{enumerate}
In practice, $X_t$ are not updated for all $t$; in these cases, we apply the Kalman smoother, $E(\hat{F}_t|\Omega_T)$, where $\Omega_T$ is all the available information in the sample, and we take into account the following two cases:

$$\hat{\Psi}_i = \left \{ \begin{matrix} \hat{\sigma}^2_i & \mbox{if } x_{it} \mbox{ is available,}
\\ \infty & \mbox{if } x_{it} \mbox{ is not available.} \end{matrix}\right. $$
Empirically, when specific data on $X_t$ are not available, \cite{HarveyPhillips1979} suggests using a diffuse value equal to $10^{7}$; however, we use $10^{32}$ according to the package \texttt{nowcast} of the R program, see \cite{deValketal2019}.

\subsubsection{Determining the number of common factors}

To detect the estimated number of common factors, $\widehat{r}$, \cite{Onatski2010} proposes a procedure when the proportion of the observed variance attributed to the factors is small relative to that attributed to the idiosyncratic term. This method determines a sharp threshold, $\delta$, which consistently separates the bounded and diverging eigenvalues of the sample covariance matrix. The author proposes the following algorithm to estimate $\delta$ and determine the number of factors:

\begin{enumerate}
\item{Obtain and sort in descending order the $N$ eigenvalues of the covariance matrix of observations, $\widehat{\Sigma}_X$}. Set $j = r_{\max}$ + 1.
\item Obtain $\widehat{\gamma}$ as the OLS estimator of the slope of a simple linear regression, with a constant of $\left\{\lambda_j, \dots ,\lambda_{j+4}\right\}$ on $\left\{(j-1)^{2/3}, \dots (j + 3)^{2/3}\right\}$, and set $\delta = 2|\widehat{\gamma}|$. 

\item Let $r_{\max}^{(N)}$ be any slowly increasing sequence (in the sense that it is $o(N)$). If $\widehat{\lambda}_k - \widehat{\lambda}_{k+1} < \delta$, set $\widehat{r} = 0$; otherwise, set $\widehat{r} = \max\lbrace k \leq r_{\max}^{(N)} \mid \widehat{\lambda}_k - \widehat{\lambda}_{k + 1} \geq \delta \rbrace$.
\item{With $j = \widehat{r}$ + 1, repeat steps 2 and 3 until convergence.}
\end{enumerate}
This algorithm is known as edge distribution, and \cite{Onatski2010} proves the consistency of $\widehat{r}$ for any fixed $\delta > 0$. \cite{corona2017determining} shows that this method works reasonably well in small samples. Two important features of this method are that the number of factors can be estimated without previously estimating the common components and that the common factors may be integrated.

%

\subsection{Nowcasting approach}

In this subsection, we describe the nowcasting approach to estimate the annual percentage variation of IGAE, denoted by $y^{*} = (y_1, \dots, y_{T^{*}})$, where $T^{*} = T-2$; hence, we focus on generating the nowcasts two steps ahead.

\subsubsection{Selecting relevant Google Trends topics} \label{google}
Currently, Google Trends topics, an up-to-date source of information that provides an index of Internet searches or queries by category and geography, are frequently used to predict economic phenomena. See, for instance, \cite{stephens2014hands} for a full review of this tool and other analytical tools from Google applied to social sciences. Other recent examples are \cite{ali2020impact}, who analyzes online job postings in the US childcare market under stay-at-home orders, \cite{goldsmith2020predicting}, who nowcast the number of workers filing unemployment insurance claims in the US, based on the intensity of search for the term ``file for unemployment'', and \cite{caperna2020googling}, who develop random forest models to nowcast country-level unemployment figures for the 27 European Union countries based on queries that best predict the unemployment rate to create a daily indicator of unemployment-related searches.

In this way, for a sample $K$ topics on Google Trends, the relevant topics $l = 0, \dots, \zeta$, with $\zeta \geqslant 0$ are selected with LASSO regression as follows:

\begin{enumerate}
\item Split the data for $t=1, \dots, T^{*}-H_g$.
\item For $h = 1$ and for the sample of size $T^{*}-H_g+h$, estimate $\widehat{\beta}^{LASSO}_{\lambda,h}$. Compute the following vector of indicator variables:

$$\widehat{\beta}_{j,h} = \left \{ \begin{matrix} 1 & \mbox{if } \widehat{\beta}^{LASSO}_{\lambda,h} \neq 0
\\ 0 & \mbox{if }  \widehat{\beta}^{LASSO}_{\lambda,h} = 0 \end{matrix}\right.$$

\item Repeat 2 until $H_g$.

\item Define the $H_g \times K$ matrix, $\widehat{\beta} = (\widehat{\beta}_1, \dots, \widehat{\beta}_K)$, where $\widehat{\beta_j} = (\widehat{\beta}_{j,1}, \dots, \widehat{\beta}_{j,H_g})'$ is an $H_g \times 1$ vector.

\item Select the $l$ significant variables that satisfy the condition $\widehat{\beta}_l = \left(\widehat{\beta}_{l \in j} | \textbf{1}\widehat{\beta}  > \varphi \right) $, where $\varphi$ is the $1-\alpha$ sample quantile of $\textbf{1}\widehat{\beta}$ with \textbf{1} being and vector $1 \times H_g$ of ones.
\end{enumerate}

With this procedure, we select the topics that frequently reduce the prediction error -- in sample -- for the IGAE estimates during the last $H_g$ months. We estimate the optimum $\lambda$ by using the \texttt{glmnet} package from the R program.

\subsubsection{Transformations}

In our case, to predict $y^{*}$, the time series $X_i = (x_{i1}, \dots, x_{iT^{*}})$ are transformed such that they satisfy the following condition:

\begin{equation} \label{eq5}
X^{*}_i = \left( f(X_i) \mid \max_{corr} (f(X_i),y^{*}) \right) .
\end{equation}
Hence, we select the $f(X_i)$ that maximizes the correlation between $y$. Consider $f(\cdot)$ as follows:

\begin{enumerate}

\item None (n)
\item Monthly percentage variation (m): $\left( \frac{X_t}{X_{t-1}}\times 100\right) - 100$
\item Annual percentage variation (a): $\left( \frac{X_t}{X_{t-12}} \times 100\right) - 100$
\item Lagged (l): $X_{t-1}$

\end{enumerate}
Note that these transformations do not have the goal of achieving stationarity, although intrinsically these transformations are stationary transformations regardless of whether $y^{*}$ is stationary; in fact, the transformations $m$ and $a$ tend to be stationary transformations when the time series are $I(1)$, which is frequent in economics; see \cite{corona2017determining}. 
Otherwise, it is necessary that $(f(X_i),y^{*})$ are cointegrated. The implications of equations (\ref{eq2}) to (\ref{eq4}) are very important because it is necessary to stationarize the system in terms that, theoretically, although some common factor, $F_t$, can be nonstationary, consistent estimates remain regardless of whether the idiosyncratic errors are stationary, see \cite{Bai2004}. 
In this way, we use the PANIC test \citep{BaiNg2004} to verify this assumption. Additionally, an alternative to estimate nonstationary common factors by using 2SM when the time series are $I(1)$ is given by \cite{Coronaetal2020}.

\subsubsection{Nowcasting approach} \label{nowcast}
Having estimated the common factors as described in subsection \ref{sec2sm} by using $X_t^{*}$ for $t = 1, \dots, T$, we estimate a linear regression model with autoregressive moving average (ARMA) errors to generate the nowcasts

\begin{equation} \label{eq6}
 y^{*}_t = a + b \tilde{F}_t + u_t \quad t = 1, \dots, T-2,
\end{equation}
where $u_t = \phi(L)u_t + \gamma(L)v_t$ with $\phi(L) = \sum_{i=1}^p \phi_i L^{i}$ and $\gamma(L) = 1 + \sum_{j=1}^q \gamma_j L^j$. The parameters are estimated by maximum likelihood. Consequently, the nowcasts are obtained by the following expression:

\begin{equation} \label{eq8}
\widehat{y}_{T^{*}+h} = \widehat{a} + \widehat{b} \tilde{F}_{T^{*}+h} +  \widehat{u}_{T^{*}+h} \quad \text{for} \quad  h = 1, 2.
\end{equation}
Note that \cite{Giannoneetal2008} propose using the model with $p = q = 0$; hence, the nowcasts are obtained by using the expression (\ref{eq8}). In our case, we estimate different models by the orders $p = 0, \dots p_{\max}$ and $q = 0, \dots q_{\max}$; thus ,the case of \cite{Giannoneetal2008} is a particular case of this expression. Now, our interest is in selecting models with similar performance for training data. In this way, we carry out the following procedure:

\begin{enumerate}
\item Start with $p = 0$ and $q = 0$.
\item Estimate the nowcasts for $T^{*}+1$ and $T^{*}+2$, namely, $\widehat{y}^{0,0} = (\widehat{y}_{T^{*}+1}, \widehat{y}_{T^{*}+2})'$.
\item Split the data for $t = 1, \dots, T^{*}-H_t.$
\item For $h = 1$ and for the sample of size $T^{*}-H_t+h$, estimate equation (\ref{eq6}), generate the nowcasts with expression (\ref{eq8}) one step ahead, and calculate the errors and absolute error (AE) as follows:
$$e^{0,0}_1 = y_{T^{*}-H_t+1} - \widehat{y}_{T^{*}-H_t+1}$$
$$ AE_1^{0,0} = | e^{0,0}|$$
\item Repeat steps 3 and 4 until $H_t$. Hence, estimate $e^{0,0} = (e^{0,0}_1, \dots, e^{0,0}_H)'$ and $AE^{0,0}= (AE^{0,0}_1, \dots, AE^{0,0}_{H_t})$. Additionally, we define the weighted AE (WAE) as $WAE^{0,0} = AE^{0,0}\Upsilon$ where $\Upsilon$ is a weighted $H_t \times 1$ matrix that penalizes the nowcasting errors such that $\Upsilon \textbf{1}' = 1.$
\item Repeat steps for all combinations of $p$ and $q$ until $p_{\max}$ and $q_{\max}$. Generate the following elements:
$$\widehat{y}(p,q) = (\widehat{y}^{0,0}, \widehat{y}^{1,0}, \dots, \widehat{y}^{p_{\max}, q_{\max}}),$$
$$e(p,q) = (e^{0,0}, e^{1,0}, \dots, e^{p_{\max}, q_{\max}}),$$
$$WAE(p,q) = (WAE^{0,0}, WAE^{1,0}, \dots, WAE^{p_{\max}, q_{\max}})',$$
where $\widehat{y}$ is a $2 \times (p_{\max} + 1)(q_{\max} + 1)$ matrix of nowcasts, $e$ is an $H_t \times (p_{\max} + 1)(q_{\max} + 1)$ matrix that contains the nowcast errors in the training data, and $WAE$ is an $H_t \times 1$ vector of the weighted errors in the training data.
\item We select the best nowcast as a function of $p$ and $q$, denoted by $\widehat{y}(p^{*},q^{*})$, where $p^{*},q^{*}$ are obtained as follows:

$$p^{*},q^{*} = \argmin \limits_{0 \leq p,q \leq p_{\max}, q_{\max}} WAE(p,q)$$
\item To use models with similar performance, we combine the nowcasts of $\widehat{y}(p^{*},q^{*})$ with models with equal forecast errors according to \cite{DieboldMariano1995} tests, by using the $e(p,q)$, carrying out pairs of tests between the model with minimum $AE(p,q)$ and the others. Consequently, from the models with statistically equal performance, we select the median of the nowcasts, namely, $\widehat{y}$.
\end{enumerate}
This nowcasting approach allows the generation of nowcasts based on a trained process, taking advantage of the information of similar models. It is clear that $\widehat{b}$ must be significant to exploit the relationship between the IGAE and the information summarized by the DFM. Note that $\Upsilon$ is a weighted matrix that penalizes the nowcasts errors. The most common form is $\Upsilon = (1/H_t, \dots, 1/H_t)'$, a $H_t \times 1$ matrix where all nowcasts errors have equal weight named in literature as mean absolute error (MAE). Therefore, we are not considering by default the traditional MAE, but rather a weighted (or equal) average of the individual AE. For example, we could have penalized with more weight the last nowcasts errors, that is, in the COVID-19 period. Also, note that we can obtain $AE(p,q)$ and estimate the median or some specific quantile for each vector of this matrix.

Note that despite root mean squared errors (RMSEs) are often used in the forecast literature, we prefer a weighted function of AEs, although in this work we use equal weights i.e., the MAE. The main advantages of MAE over RMSE are in two ways: i) it is easy to interpret since it represents the average deviation without considering their direction, while the RMSE averages the squared errors and then we apply the root, which tends to inflate larger errors and ii) RMSE does not necessarily increase with the variance of the errors. Anyway, the two criteria are in the interval $ \left[ 0, \infty \right) $ and are indistinct to the sign of errors.

\section{Data and descriptive analysis}  \label{sect4}

\subsection{Data} \label{data}

The variables to estimate the DFM are selected by using the criteria of timely and contemporaneous correlation with respect to $y^{*}$. In this sense, the model differs from the traditional literature on large DFMs, which uses a large amount of economic and financial variables; see, for example, \cite{Coronaetal2017} who use 211 time series to estimate the DFM for the Mexican case with the goal of generating forecasts for the levels of IGAE. On the other hand, \cite{Galvez2020} uses approximately 30 selected time series to generate nowcasts of Mexican quarterly GDP. Thus, our number of variables is intermediate between these two cases. However, as noted by \cite{BoivinNg2005}, in the context of DFM, we can reduce the forecast prediction error with selected variables by estimating the common components. Additionally, \cite{PoncelaRuiz2015b} and \cite{Coronaetal2020} show that with a relativity small sample size, for example, $N = 12$, we can accurately estimate a rotation of the common factors.

Consequently, given the timely and possibly contemporaneous correlation with respect to the $y^{*}$, the features of the variables considered in this work are described in Annex \ref{Anne1}.\footnote{All variables are seasonally adjusted in the following ways: i) directly downloadable from their source or ii) by applying the X-13ARIMA-SEATS.}

Hence, we initialized with 68 time series divided into three blocks. The first block is timely traditional information such as the Industrial Production Index values for Mexico and the United States, business confidence, and exports and imports, among many others. In this block, all variables are monthly. In the second block, we have the high-frequency traditional variables such as Mexican stock market index, nominal exchange rate, interest rate and the Standard Poor's 500. These variables can be obtained daily, but we decide to use the averages to obtain monthly time series. Finally, for the high-frequency nontraditional variables, we have daily variables such as the social media mobility index obtained from Twitter and the topics extracted from Google Trends. These topics are manually selected according to several phenomena that occur in Mexican society, such as politicians' names, natural disasters, economic themes and topics related to COVID-19, such as coronavirus, quarantine, or facemask. The Google Trends variable takes a value of 0 when the topic is not searched in the time span and 100 when the topic has the maximum search in the time span. 
In a similar way, although these variables are expressed as monthly variables, for the social media mobility index, we average the daily values, and for Google Trends we download the variables by month. 

The social media mobility index is calculated based on Twitter information. We select around  70,000 daily tweets georeferenced to the Mexican, each one is associated with a bounding box.  Then, movement data analysis is performed by identifying users and their sequence of daily tweets: a trip is considered for each pair of consecutive geo-tagged tweets found in different bounding boxes. The total number of trips per day is obtained and divided by the average number of users in the month. The number obtained can be interpreted as the average number of trips that tweeters make per day.

To select the relevant topics, we apply the methodology described in subsection \ref{google} by using $H_g = 36$ and $\alpha = 0.10$; consequently, we select the topics that are relevant in 90\% of cases in the training data. In this way, the significant topics are quarantine and facemask. 

\begin{figure}[H]
\begin{center}
\includegraphics[scale = 0.3]
{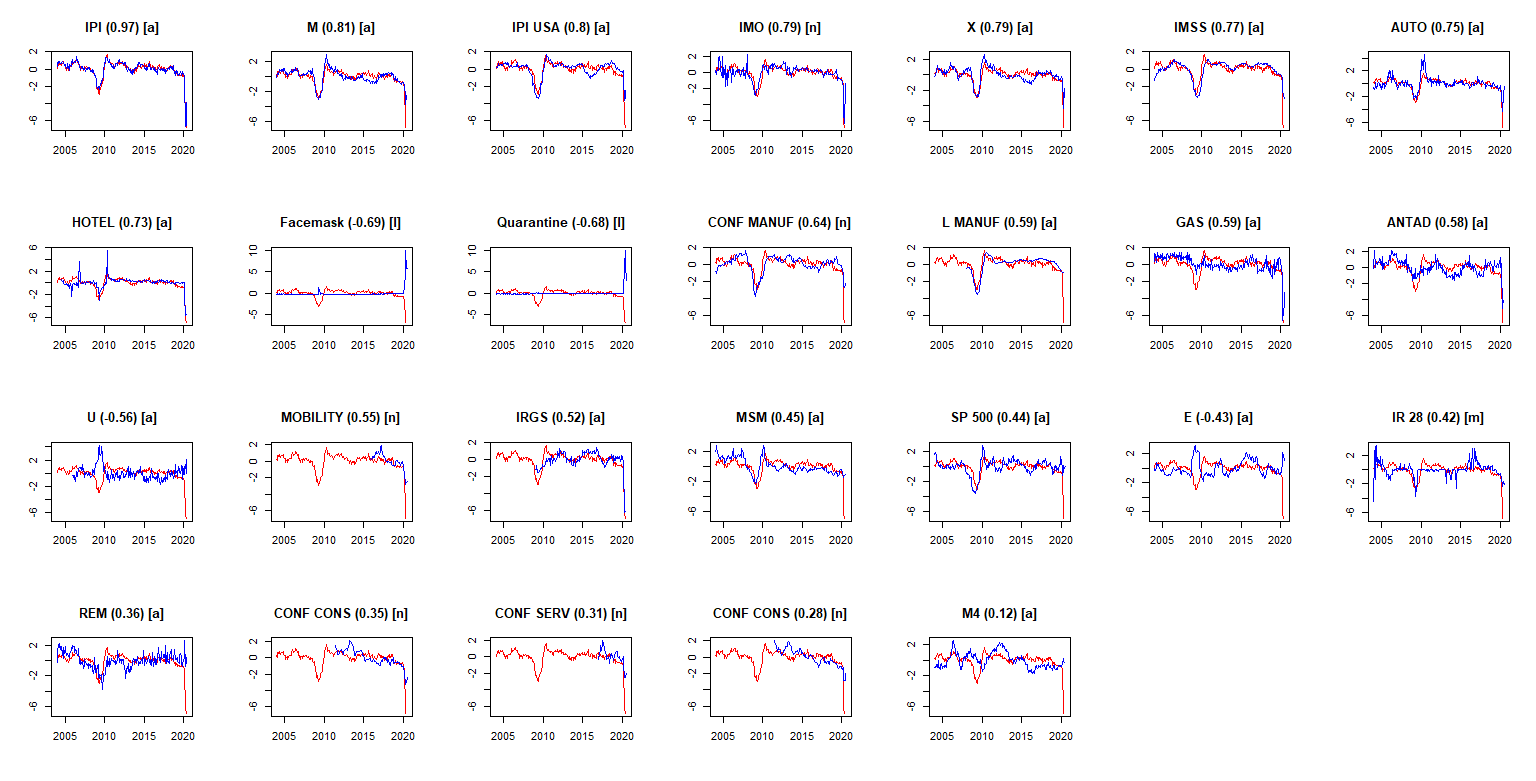}
\end{center}
\caption{Blue indicates the specific $X_i^{*}$, and red indicates the specified $y^{*}$. Numbers in parentheses indicate the linear correlation and those between brackets the transformation.}
\label{FigVar}
\end{figure}

Once $X$ is defined, we apply the transformations suggested by equation (\ref{eq5}) to define $X^{*}$. Figure \ref{FigVar} shows each $X_i^{*}$ ordered according to its correlation with $y^{*}$.

We can see the behavior of each variable, and industrial production is the variable with the most correlated time series with the IGAE, followed by imports and industrial production in the United States. Note that nontraditional time series are also correlated with $y^{*}$ such as facemask, quarantine and the social mobility index. Finally, the variables less related to the IGAE are the variables related to business confidence and the monetary aggregate M4.

\begin{figure}[H]
\begin{center}
\includegraphics[scale = 0.5]
{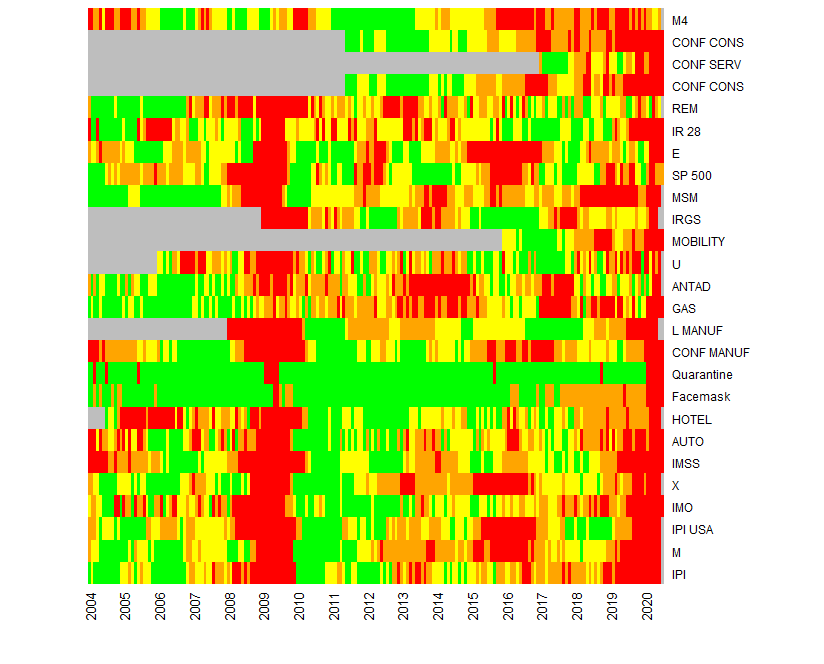}
\end{center}
\caption{Heat map plot of the variables. The time series inversely related to the IGAE are converted to have a positive relationship with it. We estimate the empirical quantiles $\varphi(\cdot)$ according to their historical values. The first quantile $(\varphi(X_i^{*}) < 0.25)$ is in red, the second quantile $(0.25 < \varphi(X_i^{*}) < 0.50)$ is in orange, the third quantile $(0.50 < \varphi(X_i^{*}) < 0.75)$ is in yellow, and finally, the fourth quantile $(0.75 < \varphi(X_i^{*}))$ is green. Gray indicates that information is not available.}
\label{FigHea}
\end{figure}

To summarize whether the time series capture prominent macroeconomic events as the 2009 financial crisis and the economic deceleration in effect since 2019, Figure \ref{FigHea} shows the heat map by time series plotted in Figure \ref{FigVar}

We can see that during the 2009 financial crisis, the variables are mainly red, including the Google Trends variables, which is reasonable because the AH1N1 pandemic also occurred during March and April of 2009. Additionally, during 2016, some variables related to the international market were red, for example, the US industrial production index, the exchange rate and the S\&P 500. Note that since 2019, all variables are orange or red, denoting the weakening of the economy. Consequently, it is unsurprising that the estimated common factor summarizes these dynamics. Note that this graph has only a descriptive objective. It cannot be employed to generate recommendations for policy making because that some variables may be nonstationary.

\subsection{Timely estimates}
\begin{figure}[H]
\begin{center}
\includegraphics[scale = 0.7]
{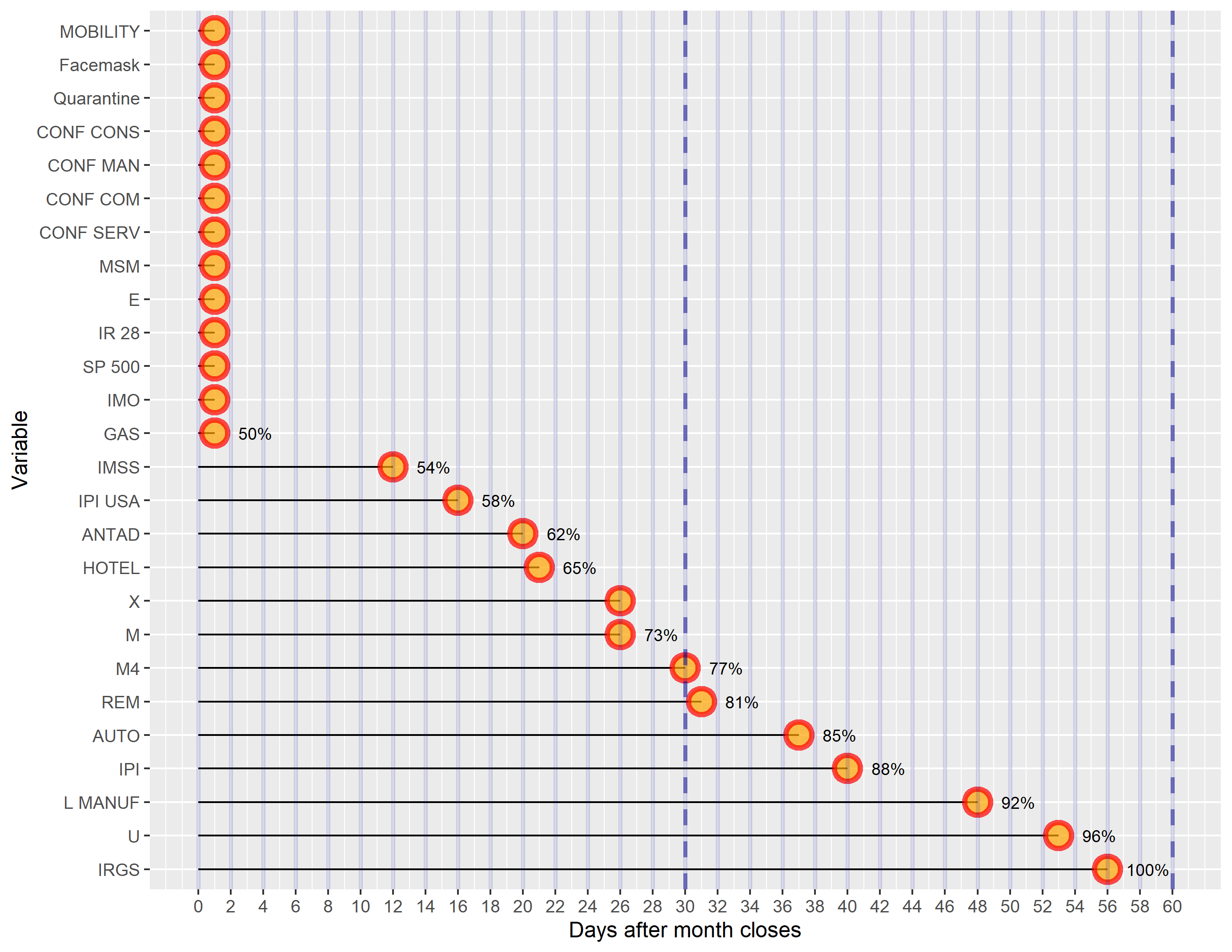}
\end{center}
\caption{Percentage of updated information to carry out the nowcasts $T^{*} + 2$ once the current month $T^{*}$ is closed.}
\label{FigMat}
\end{figure}

The nowcasts depend on the dates of the information released. Depending on the day of the current month, we can obtain nowcasts with a larger or smaller percentage of updated variables. For example, it is clear that the high-frequency variables are available in real time, but the traditional and monthly time series, with are timely with respect to the IGAE, are available on different dates according to the official release dates. Figure \ref{FigMat} shows the approximate day when the information is released for $T^{*} + 2$ after the current month $T^{*}$.

We can see that traditional and nontraditional high-frequency variables, business confidence and fuel demand, can be obtained on the day after the month $T^{*}$ is closed. 
This indicates that on the first day of month $T^{*} + 1$, we can generate the nowcasts to $T^{*} + 2$ with approximately 50\% of the updated information and 81\% for the current month, $T^{*} + 1$. Note that on day 12, the IMSS variable is updated, and on day 16, the IPI USA is updated. These variables are highly correlated with $\widehat{y}$ with linear correlations of 0.77 and 0.80, respectively. Consequently, in official statistics, we recommend conducting the nowcasts on the first day of $T^{*} + 1$ and 16 days after, updating the nowcasts with two timely traditional and important time series, taking into account the timely estimates but with relevant variables updated.\footnote{Note that IPI represents around the 34\% of the monthly GDP, and represents more than 97\% of the second grand economic activity. Given that the IPI is updated around 10 days after the end of the reference month, this information is very valuable to carry out the $T^{*} + 1$ nowcasts.}

In this work, the update of the database is August 13, 2020; consequently, we generate the nowcasts 13 days before the official result of June 2020 and 43 days before the official value of July 2020, having 88\% and 52\% of updated variables at $T^{*} + 1$ and $T^{*} + 2$, respectively.

\section{Nowcasting results} \label{sect5}

\subsection{Estimating the common factors and the loading weights}
\begin{figure}[H]
\begin{center}
\includegraphics[scale = 0.6]
{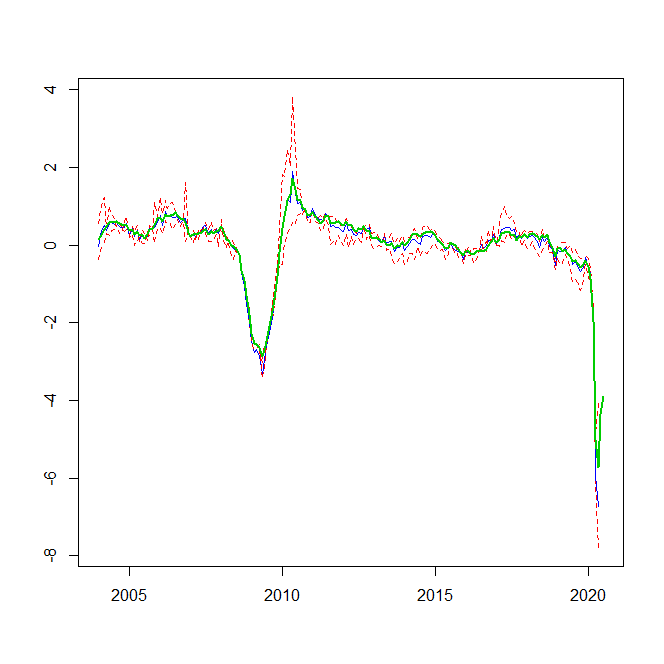}
\end{center}
\caption{Factor estimates. The blue line is the static common factor, the red lines are their confidence intervals, and the green line is the smoothed or dynamic common factor.}
\label{FigFac}
\end{figure}

By applying the \cite{Onatski2010} procedure to the covariance matrix of $X^{*}$, we can conclude that $\hat{r} = 1$ is adequate to define the number of common factors. Hence, the estimated static common factor obtained by PCs by using the set of variables, $X^{*}$, their confidence intervals at 95\%, and the dynamic factor estimates by applying the 2SM procedure with $k = 1$ lags, are presented in Figure \ref{FigFac}

\begin{figure}[H]
\begin{center}
\includegraphics[scale = 0.6]
{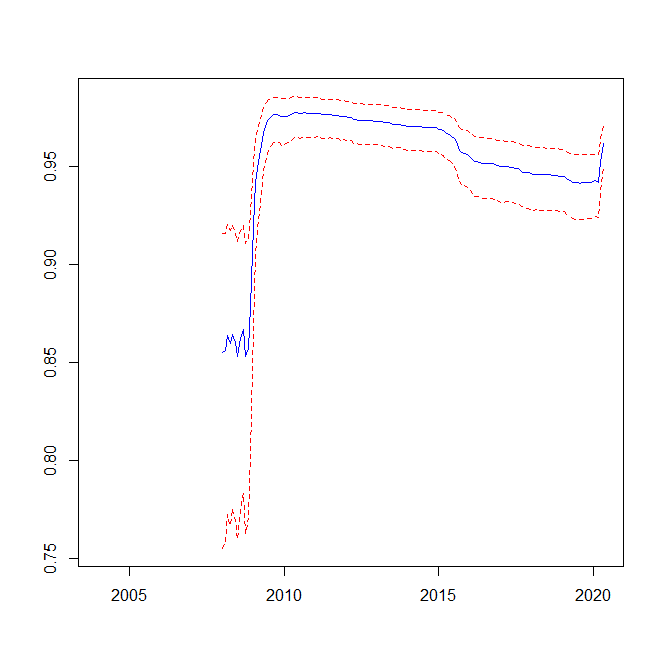}
\end{center}
\caption{Blue line is $Corr(\tilde{F}_t, y^{*})$ from January 2008 to May 2020. Red lines represent the confidence interval at 95\%.}
\label{FigCor}
\end{figure}

We observe the common factors summarizing the previous elements representing the decline in the economy in 2009 and 2020. Note that in the last period, the dynamic common factor shows a slight recovery of the economy because this common factor supplies more timely information than the static common factor. Thus, the static common factor has information until May 2020, while the dynamic factor has information until July 2020. Note that the confidence intervals are closed with respect to the static common factor, which implies that the uncertainty attributed to the estimation is well modeled. It is important to analyze the contemporaneous correlation with respect to IGAE. Thus, Figure \ref{FigCor} shows the correlation coefficient of $\tilde{F}_t$ with $y^{*}$ since 2008.

We see that the correlation is approximately 0.86 prior to the financial crisis of 2009, increasing from this year to 0.98, showing a slight decrease since 2011, dropping in 2016 to 0.95 and fully reaching levels of 0.96 since 2020. The confidence intervals are between 0.75 and 0.97 during all sample being the smallest value during the first years of the sample and the largest one in the final of period. Consequently, we can exploit the contemporaneous relationship between the dynamic factor and the IGAE to generate their nowcasts for the two following months that the common factors have estimated with respect to the IGAE.

\begin{figure}[H]
\begin{center}
\includegraphics[scale = 0.6]
{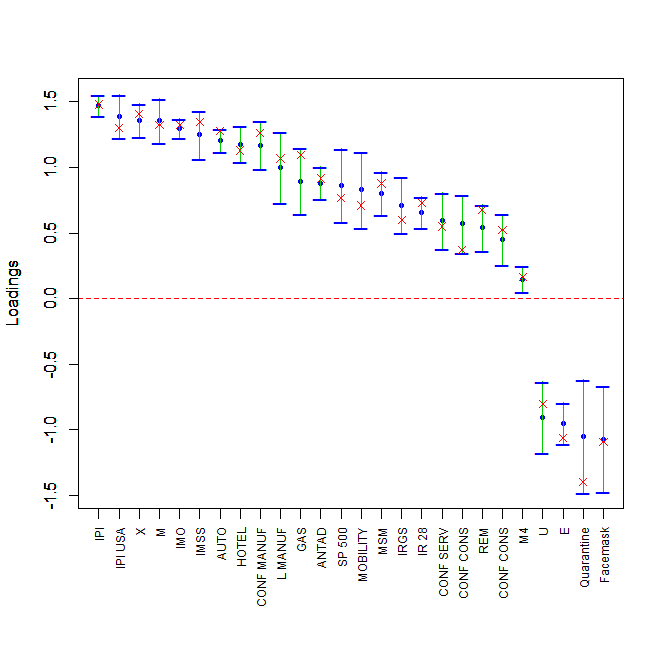}
\end{center}
\caption{Factor loadings. The blue point is each $\hat{P}_i$ with its respective 95\% confidence interval. Red curves are the $\tilde{P}_i$.}
\label{FigLoad}
\end{figure}

Having estimated the dynamic factor by the 2SM approach, we show the results of the loading weight estimates that capture the specific contribution of the common factor to each variable, or in other words, given the PC restrictions, they can be seen as $N$ times the contribution of each variable in the common factor. We compute the confidence interval at 95\% denoted by $CI_{\hat{P}, 0.05}$. Once the dynamic factor is estimated by using the Kalman smoother, it is necessary to reestimate the factor loadings to have $\hat{P} = f(\tilde{F})$, such that $\tilde{F} = g(\tilde{P})$. To do so, we use Monte Carlo estimation iterating 1,000 samples and select the replication that best satisfies the following condition:

$$\tilde{F} \approx  X\tilde{P}/N \quad \mbox{s.t} \quad \tilde{P} \in CI_{\hat{P}, 0.05}.$$
The results of the estimated factor loadings are shown in Figure \ref{FigLoad}. The loadings are ordered from the most positive contribution to the most negative.

We observe several similarities with respect to Figure \ref{FigVar}. Note that the more important variables in the factor estimates are the industrial production of Mexico and the U.S., exports and imports along with Google Trends topics such as quarantine and facemask, which makes sense in the COVID-19 period. Obviously, when these variables are updated, it will be more important to update the nowcasts. In this way, note that Google Trends are available in real time. Other timely variables, such as IMO, CONF MANUF, GAS, S\&P 500, MOBILITY and E, are also very relevant. However, note that all variables are significant in all cases, and the confidence interval does not contain zero. The less important variables are M4, the business confidence of the construction sector and remittances. 
Also, note that the most relevant variables are very timely with respect to the IGAE: the industrial production index of Mexico and the U.S. are updated around days 10 and 16 for $T^{*}+1$ and $T^{*} + 2$, respectively, once closed the current month; furthermore, the exports and imports are updated for $T^{*} + 2$ by 25th day, while IMO and IMSS are updated since the first day and 12th day, respectively for $T^{*} + 2$. Consequently, this allows us to have more accurate and correlated estimates since the first day of the current month for both, $T^{*}+1$ and $T^{*} + 2$.

As we have previously noted, to obtain a consistent estimation of $\tilde{F}$ and $\hat{P}$ it is necessary that $\hat{\varepsilon}$ be stationary. We check this point with the PANIC test of \cite{BaiNg2004}, concluding that we achieved stationarity in the idiosyncratic component, obtaining a statistic of 6.6 that generates a p-value of 0.00; hence, $\hat{\varepsilon}$ does not have a unit root. Additionally, we can verify with the augmented Dickey-Fuller test that $\tilde{F}$ is stationary with a p-value of 0.026; consequently, we also achieved stationarity in $X^{*}$.

\subsection{Nowcasts in data test}
\begin{figure}[H]
\begin{center}
\includegraphics[scale = 0.4]
{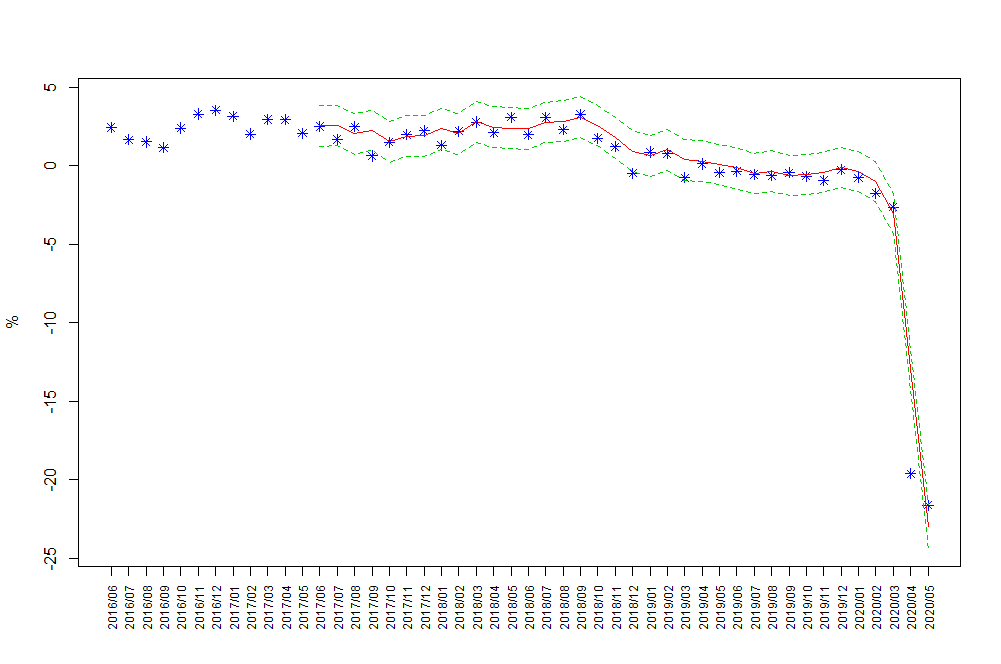}
\end{center}
\caption{Nowcasts of training model. Asterisks are the observed values, the red line depicts the nowcasts, and the green lines are the confidence intervals.}
\label{FigPer}
\end{figure}

We apply the procedure described in subsection \ref{nowcast} by using a  $\Upsilon = (1/H_t, 1/H_t, \dots, 1/H_t)'$; then, we assume that each AE has equal weight over time in step 5. 
Additionally, we fix $p_{\max} = q_{\max} = 4$. The obtained results indicate that the optimums $p^{*}$ and $q^{*}$ are selected to be equal to 4. Consequently, the best model is the following:

\begin{equation} \label{eqDFM}
\begin{split}
y^{*}_{t} = \underset{(0.1553)}{1.7160}+\underset{(0.0393)}{1.2625} \tilde{F}_{t} + \underset{(0.0830)}{0.3664} \widehat{u}_{t-1} +  \underset{(0.0932)}{1.0741} \widehat{u}_{t-2} + \underset{(0.0912)}{0.2794} \widehat{u}_{t-3} \underset{(0.0811)}{-0.8036} \widehat{u}_{t-4} +  \\ 
\underset{(0.0893)}{0.1001} \widehat{v}_{t-1}  \underset{(0.098)}{-0.949} \widehat{v}_{t-2} \underset{(0.1002)}{-0.4736} \widehat{v}_{t-3} + \underset{(0.0756)}{0.5904} \widehat{v}_{t-4} + \widehat{v}_{t} \quad \hat{\sigma}^{2} = 0.4763.
\end{split}
\end{equation}
Note that all coefficients are significant and the contribution of the factor over the IGAE is positive. Additionally, estimating the Ljung-Box test over the residuals produces a result of serially uncorrelated. This model generates the following historical nowcasts one step ahead during $H_t = 36$ months that are presented in Figure \ref{FigPer}

We can see that the nowcast model performs well given that in 92\% of cases, the observed values are within the confidence interval at 95\%. The MAE (equal weights in $\Upsilon$) is 0.65, and the mean absolute annual growth of IGAE is 2.55\%. Regarding the median of the AEs, the estimated value is 0.36. These statistics are very competitive with respect to the model estimated by Statistics Netherlands, see \cite{Kuiper2020Now}. They also estimate common factors to generate the nowcasts of the annual variation of quarterly Netherlands GDP. According to Table 7.2 in their work, the root of the mean of squared forecast errors is between 0.91 and 0.67 during 2011 and 2017. Additionally, the confidence interval captures approximately 70\% of the observed values. Therefore, our nowcast approach generates good results even when considering a monthly variable and COVID-19. 

In addition, we compare our results to \cite{Coronaetal2017}, which forecasts IGAE levels two steps ahead. To have comparable results between such study and this one, we take the median of the root squared errors obtained by the former just for the first step forward, which is between 0.4 and 0.5, while the current work generates a median AEs of 0.397 for the last $H_t = 36$ months, including the COVID-19 period. Therefore, our approach is slightly more accurate when nowcasting the IGAE levels. Note that the number of the variables is drastically less, 211 there versus 68 here. Another nowcasting model to compare with is INEGI's ``Early Monthly Estimation of Mexico's Manufacturing Production Level,"\footnote{\url{https://www.inegi.org.mx/investigacion/imoam/}} whose target variable is manufacturing activity, generating the one step ahead nowcasts by using a timely electricity indicator. The average MAE for the annual percentage variation of manufacturing activity in the last 36 months, from August 2017 to July 2020, is 1.35. Consequently, in a similar sample period, we have a smaller average MAE than another nowcasting model where its monthly target variable is specified as annual percentage variation.

In order to contrast the results of our approach with those obtained by other procedures, we consider the following two alternative models:

\begin{itemize}

\item Naive model: We assume that all variables have equal weights in the factor, consequently, we standardize the variables used in the DFM, $X_t^{*}$, and by averaging their rows, we obtain a $F_t^{*}$. Then, we use this naive factor in a linear regression in order to obtain the nowcasts by the last $H_t = 36$ months.

\item DFM without nontraditional information: We estimate a traditional DFM similar to \cite{Coronaetal2017} or \cite{Galvez2020}, but using only economic and financial time series, i.e. without considering the social mobility index and the relevant topics extracted from Google Trends. Hence, we carry out the last $H_t = 36$ nowcasts.
\end{itemize}

\begin{figure}[H]
\begin{center}
\includegraphics[scale = 0.5]
{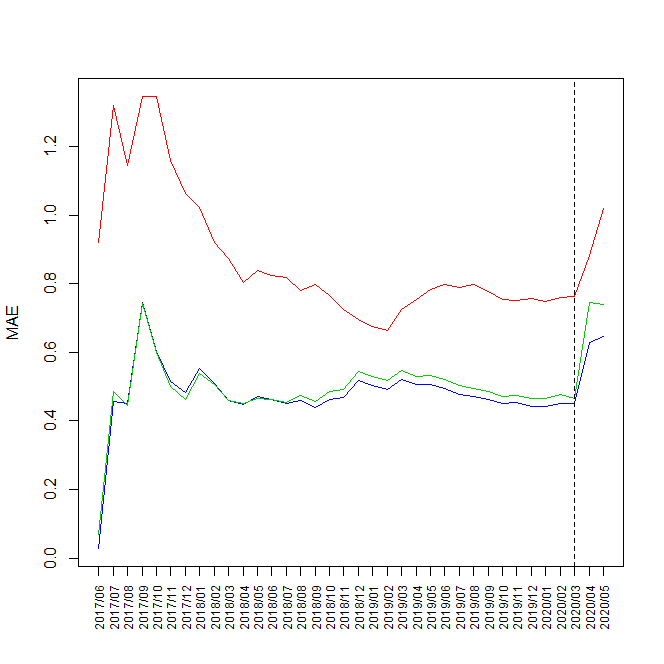}
\end{center}
\caption{Cummulative MAEs for models in training data. Blue is the nowcasting approach suggested in this work, red is the naive model, green is the traditional DFM (without nontraditional information). The vertical line indicates indicates the COVID-19 onset period.}
\label{RMSE}
\end{figure}

Figure \ref{RMSE} shows the accumulated MAEs for the training period by the previous two models and the obtained by equation (\ref{eqDFM}).

We can see that, in training data the named naive model is the one with the weakest performance, followed by traditional DFM. Specifically, the MAE is 1.02 for the naive model, 0.74 when using DFM without nontraditional information and, as we have commented, 0.65 for the incumbent model, which includes this type of information. Note that the use of nontraditional information does not affects the behaviour of the MAEs previous to COVID-19 pandemic and reduces the error during this period. Consequently, the performance of the suggested approach is highly competitive when compared with i) similar models for nowcasting of GDP, ii) models that estimate the levels of the objective variable and iii) alternative models that can be used in practice.

\subsection{Current nowcasts}
\begin{figure}[H]
\begin{center}
\includegraphics[scale = 0.5]
{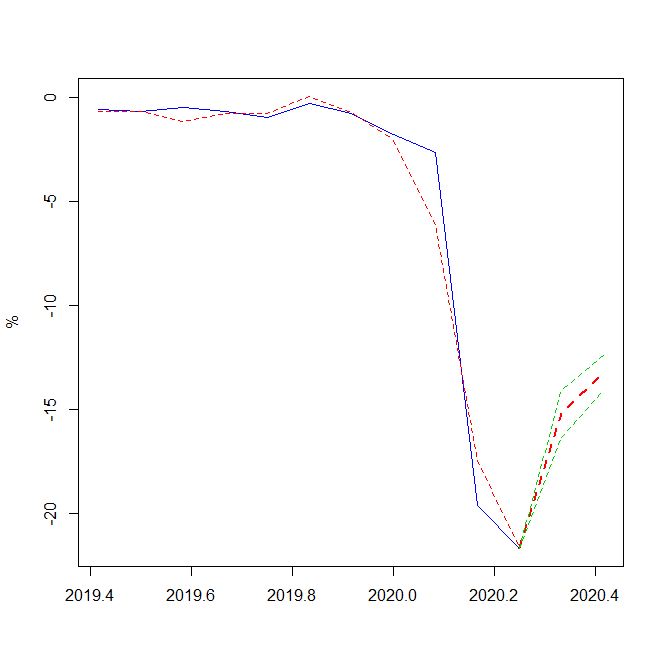}
\end{center}
\caption{Nowcasts for June and July 2020. The blue line indicates the observed values, the red small dotted line the fit of the model, the red large dotted line the nowcasts and the green lines the confidence intervals.}
\label{FigNow}
\end{figure}

Having verified our approach in the previous section as highly competitive to capture the real observed values, the final nowcasts for the IGAE annual percentage  variation for June and July 2020 are shown in Figure \ref{FigNow}. These are obtained after combining the statistically equal models to the best model with the approach previously described and the traditional nowcasting model of \cite{Giannoneetal2008}, weighting both nowcasts according to their MAEs.\footnote{Note that the model of \cite{Giannoneetal2008} uses only the estimated dynamic factors as regressors, i.e., linear regression models. Our approach also considers the possibility to model the errors with ARMA models. In order to consider nowcasts associated to specifically the dynamic of the common factors, we take into account the \cite{Giannoneetal2008} model although its contribution in the final nowcasts is small given that, frequently, during the test period, the nowcast errors are greater than the regression models with ARMA errors.}

We expect a slight recovery of the economy in June and July 2020, obtaining nowcasts of -15.2\% and -13.2\%, respectively, with confidence intervals of (-16.3, -14.1) and (-14.1, -12.4) for both months. Considering the observed values for June, released on August 25 by INEGI, the annual percentage change for the IGAE was -14.5\%; consequently, the model is very accurate since the deviation from the real value was 0.7\% and falls within the confidence interval.

\subsection{Historical behaviour of nowcasting model in real time: COVID-19}

The procedure described in the previous subsection allows to generate nowcasts using databases with different cut dates. In this way, we carry out the procedure updating the databases twice a month during the COVID-19 period. Table \ref{Tab1} summarizes the nowcasts results, comparing them with the observed values.

\begin{table}[H]
\scriptsize
  \centering
  \caption{Nowcasts with different updates in COVID-19 times: annual percentage variation of IGAE}
    \begin{tabular}{lr|cccccc} \hline \hline
          &       & \multicolumn{5}{c}{\textbf{Date of nowcasts}} \\ \hline
    \textbf{Date}  & \textbf{Observed} & \multicolumn{1}{r}{\textbf{04/06/2020}} & \multicolumn{1}{r}{\textbf{18/06/2020}} & \multicolumn{1}{r}{\textbf{07/07/2020}} & \multicolumn{1}{r}{\textbf{16/07/2020}} & \multicolumn{1}{r}{\textbf{06/08/2020}} & \multicolumn{1}{r}{\textbf{12/08/2020}} \\ \hline
    2020/04 & \textbf{-19.7} & -18.3 & -18.0 &       &       & & \\
    2020/05 & \textbf{-21.6} & -20.4 & -21.0 & -21.8 & -20.4 &  &\\
    2020/06 & \textbf{-14.5} &       &       & -16.6 & -16.4 & -15.5 & -15.2 \\
    2020/07 &       &       &       &       &       & -13.9 & -13.2\\ \hline
    \end{tabular}%
  \label{Tab1}%
\end{table}%
We can see that in June 4, 2020, the nowcasts were very accurate, capturing the drastic drop occurred in April (previous month was -2.5\%) and May, with absolute discrepancies of 1.4 and 1.2\% respectively. The update of June 18, 2020 shows a slight accuracy improvement. The following two nowcasts generate also closes estimates with respect to the observed value of May, being the more accurate, the updated carried out in July 7, 2020. Note the the last updates generate nowcasts by June around -16.6 and -15.2\%, being the more accurate the last nowcasts described in this work, with an absolute error of 0.7\%. Considering these results, our approach anticipates the drop attributed to the COVID-19 and foresees and slight recovery since June, although it is also weak. According to \cite{Galvez2020}, the IGAE's accurate and timely estimates can drastically improve the nowcasts of the quarterly GDP; consequently, the benefits of our approach are also related to quarterly time series nowcast models.

\section{Conclusions and further research}\label{sect6}
In this paper, we contribute to the nowcasting literature by focusing on the two step-ahead of the annual percentage variation of IGAE, the equivalently of the Mexican monthly GDP, during COVID-19 times. For this purpose, we use statistical and econometric tools to obtain accurate and timely estimates, even, around 50 days before that the official data. The suggested approach consists in using LASSO regression to select the relevant topics that affect the IGAE in the short term, build a correlated and timely database to exploit the correlation among the variables and the IGAE, estimate a dynamic factor by using the 2SM approach, training a linear regression with ARMA errors to select the better models and generate current nowcasts.

We highlight the following key results. We can see that our approach is highly competitive considering other models as naive regressions or traditional DFM, our procedure frequently captures the observed value, both, in data test and in real time, obtaining absolute errors between 0.2\% and 1.4\% during the COVID-19 period. Another contribution of this paper lies in a statistical point of view, given that we compute the confidence interval of the factor loadings and the factor estimates, verifying the significance of the factor on each variable and the uncertainty attributed to the factor estimates. Additionally, we consider some econometric issues to guarantee the consistency of estimates like stationarity in idiosyncratic noises and uncorrelated errors in nowcasting models. Additionally, it is of interest to denote in-sample performance whether the nowcast error increases when using monthly  versus quarterly data.

Future research topics emerged when doing this research. One is the implementation of an algorithm to allow to estimate nonstationary common factors and making the selection to the number of factors flexible, such as the one developed in \cite{Coronaetal2020}, to minimize a measure of nowcasting errors. Another interesting research line is to incorporate machine learning techniques to automatically select the possible relevant topics from Google Trends. Also, it would be interesting to incorporate IPI information as restrictions to the nowcasts, by exploring some techniques to incorporate nowcasts restrictions when official countable information is available. Finally, for future research in this area, its worth to deep into the effects of monthly timely estimate variables versus quarterly time series in nowcasting models, this can be achieved by Monte Carlo analysis with different data generating process which can occur in practice to compare the increase in the error estimation when distinct frequencies of time series are used.

\section*{Acknowledgements}
The authors thankfully acknowledge the comments and suggestions carried out by the authorities of INEGI Julio Santaella, Sergio Carrera and Gerardo Leyva. The seminars and meetings organized by them were very useful to improve this research. To Elio Villasen\~or who provided the Twitter social mobility index and Manuel Lecuanda by the discussion about the Google Trend topics to be considered. Partial financial support from CONACYT CB-2015-25996 is gratefully acknowledged by Francisco Corona and Graciela Gonz\'alez-Far\'ias.

\addcontentsline{toc}{section}{References} 
\bibliographystyle{apalike}
\bibliography{References}

\newpage

\section*{Annexes}
\setcounter{table}{0}
\renewcommand\tablename{Annex}

\begin{table}[htbp]
  \tiny  
  \centering
  \caption{Database} 
    \begin{tabular}{cccc} \hline \hline
    \multicolumn{4}{c}{\textbf{Traditional and timely information}} \\ \hline
    \textbf{Short} & \textbf{Variable} & \textbf{Source} & \textbf{Time Span} \\ \hline
    ANTAD & Total sales of departmental stores & ANTAD & 2004/01-2020/06 \\
    AUTO  & Automobiles production & INEGI & 2004/01-2020/07 \\
    CONF COM & Right time to invest (Commerce) & INEGI & 2011/06-2020/07 \\
    CONF CONS & Right time to invest (Construction) & INEGI & 2011/06-2020/07 \\
    CONF MANU & Right time to invest (Manufacturing) & INEGI & 2004/01-2020/07 \\
    CONF SERV & Right time to invest (Services) & INEGI & 2017/01-2020/07 \\
    GAS   & Fuel demand & SENER & 2004/01-2020/07 \\
    HOTEL & Hotel occupancy & Tourism secretariat & 2004/01-2020/06 \\
    IMO   & Index of manufacturing orders & INEGI & 2004/01-2020/07 \\
    IMSS  & Permanent and eventual insureds of the Social Security  & IMSS  & 2004/01-2020/07 \\
    IPI   & Industrial Production Index & INEGI & 2004/01-2020/06 \\
    IPI USA & Industrial Production Index (USA) & BEA   & 2004/01-2020/07 \\
    IRGS  & Income of retail goods and services & INEGI & 2008/01-2020/05 \\
    L MANUF & Trend of labor in manufacturing & INEGI & 2007/01-2020/05 \\
    M     & Total imports & INEGI & 2004/01-2020/06 \\
    M4    & Monetary aggregate M4 & Banxico & 2004/01-2020/06 \\
    REM   & Total remittances  & Banxico & 2004/01-2020/06 \\
    U     & Unemployment rate & INEGI & 2005/01-2020/06 \\
    X     & Total exports & INEGI & 2004/01-2020/06 \\ \hline
    \multicolumn{4}{c}{\textbf{High frequency traditional variables}} \\ \hline
    \textbf{Short} & \textbf{Variable} & \textbf{Source} & \textbf{Time Span} \\
    E     & Nominal exchange rate & Banxico & 2004/01-2020/07 \\
    IR 28 & Interest rate (28 days) & Banxico & 2004/01-2020/07 \\
    MSM   & Mexican stock market index & Banxico & 2004/01-2020/07 \\
    SP 500 & Standard \& Poor's 500  & Yahoo! finance & 2004/01-2020/07 \\ \hline
    \multicolumn{4}{c}{\textbf{High frequency nontraditional variables}} \\ \hline
    \textbf{Short} & \textbf{Variable} & \textbf{Source} & \textbf{Time Span}
     \\
    AH1N1 & AH1N1 online search index & Google & 2004/01-2020/07 \\
    AMLO  & AMLO online search index & Google & 2004/01-2020/07 \\
    Ayotzinapa & Ayotzinapa online search index & Google & 2004/01-2020/07 \\
    Calder\'on & Calder\'on online search index & Google & 2004/01-2020/07 \\
    C\'artel & C\'artel online search index & Google & 2004/01-2020/07 \\
    Casa Blanca & Casa Blanca online search index & Google & 2004/01-2020/07 \\
    Chapo & Chapo online search index & Google & 2004/01-2020/07 \\
    China & China online search index & Google & 2004/01-2020/07 \\
    Coronavirus & Coronavirus online search index & Google & 2004/01-2020/07 \\
    Corrupci\'on & Corrupci\'on online search index & Google & 2004/01-2020/07 \\
    Crisis econ\'omica & Crisis econ\'omica online search index & Google & 2004/01-2020/07 \\
    Crisis sanitaria & Crisis sanitaria online search index & Google & 2004/01-2020/07 \\
    Cuarentena & Cuarentena online search index & Google & 2004/01-2020/07 \\
    Cubrebocas & Cubrebocas online search index & Google & 2004/01-2020/07 \\
    Desempleo & Desempleo online search index & Google & 2004/01-2020/07 \\
    D\'olar & D\'olar online search index & Google & 2004/01-2020/07 \\
    Elecciones & Elecciones online search index & Google & 2004/01-2020/07 \\
    EPN   & EPN online search index & Google & 2004/01-2020/07 \\
    Gasolina & Gasolina online search index & Google & 2004/01-2020/07 \\
    Homicidios & Homicidios online search index & Google & 2004/01-2020/07 \\
    Huachicol & Huachicol online search index & Google & 2004/01-2020/07 \\
    Inflaci\'on & Inflaci\'on online search index & Google & 2004/01-2020/07 \\
    Inseguridad & Inseguridad online search index & Google & 2004/01-2020/07 \\
    Mascarilla N95 & Mascarilla N95 online search index & Google & 2004/01-2020/07 \\
    Medidas econ\'omicas & Medidas econ\'omicas online search index & Google & 2004/01-2020/07 \\
    Migraci\'on & Migraci\'on online search index & Google & 2004/01-2020/07 \\
    Migrantes & Migrantes online search index & Google & 2004/01-2020/07 \\
    MOBILITY & Media mobility index & Twitter & 2004/01-2020/07 \\
    Morena & Morena online search index & Google & 2004/01-2020/07 \\
    Muertos & Muertos online search index & Google & 2004/01-2020/07 \\
    Muro  & Muro online search index & Google & 2004/01-2020/07 \\
    Pacto & Pacto online search index & Google & 2004/01-2020/07 \\
    PAN   & PAN online search index & Google & 2004/01-2020/07 \\
    Pandemia & Pandemia online search index & Google & 2004/01-2020/07 \\
    PEMEX & PEMEX online search index & Google & 2004/01-2020/07 \\
    Peso  & Peso online search index & Google & 2004/01-2020/07 \\
    Petr\'oleo & Petr\'oleo online search index & Google & 2004/01-2020/07 \\
    PRI   & PRI online search index & Google & 2004/01-2020/07 \\
    Recesi\'on & Recesi\'on online search index & Google & 2004/01-2020/07 \\
    Reformas & Reformas online search index & Google & 2004/01-2020/07 \\
    Salario & Salario online search index & Google & 2004/01-2020/07 \\
    Sismo & Sismo online search index & Google & 2004/01-2020/07 \\
    Tipo de cambio & Tipo de cambio online search index & Google & 2004/01-2020/07 \\
    Trump & Trump online search index & Google & 2004/01-2020/07 \\
    Violencia & Violencia online search index & Google & 2004/01-2020/07 \\ \hline
    \end{tabular}%
  \label{Anne1}%
\end{table}%

\end{document}